**Wavelength-Dependent Photoluminescence Mapping of Depth-Dependent Degradation Under Thermal Stress in Phenethylammonium Tetrafluoroborate Treated Perovskite Solar Cells**

Julisa Juarez[1], Akash Dasgupta[1], Alan Zhan[1], Lamtiur P. Samosir[1], David S. Ginger[1*]

1. Department of Chemistry, University of Washington, Seattle, WA 98195, USA

*Corresponding author: dginger@uw.edu

**Abstract**

Studying real-time degradation in perovskite semiconductors remains an important challenge. Conventional methods often cannot separate interfacial and bulk processes or rely on destructive analysis. Here, we demonstrate multiwavelength excitation photoluminescence (PL) as a nondestructive approach to distinguish interface- and bulk-dominated recombination losses. Using phenethylammonium tetrafluoroborate ($PEABF_4$) treatment as a model, we use PL to show that $PEABF_4$ passivates the electron transport layer (ETL)/perovskite interface, reducing interfacial recombination losses from fullerene deposition. However, upon thermal aging, $PEABF_4$-treated stacks degrade more rapidly, with recombination losses broadening from the ETL/perovskite interface into the absorber. Diffusion–recombination simulations support interpreting excitation-wavelength-dependent PL decay signatures consistent with changes in surface recombination velocity and bulk lifetime, providing a framework for depth-sensitive comparisons during aging. Collectively, these results further establish multiwavelength excitation PL as a platform for studying recombination losses during aging of perovskite films.

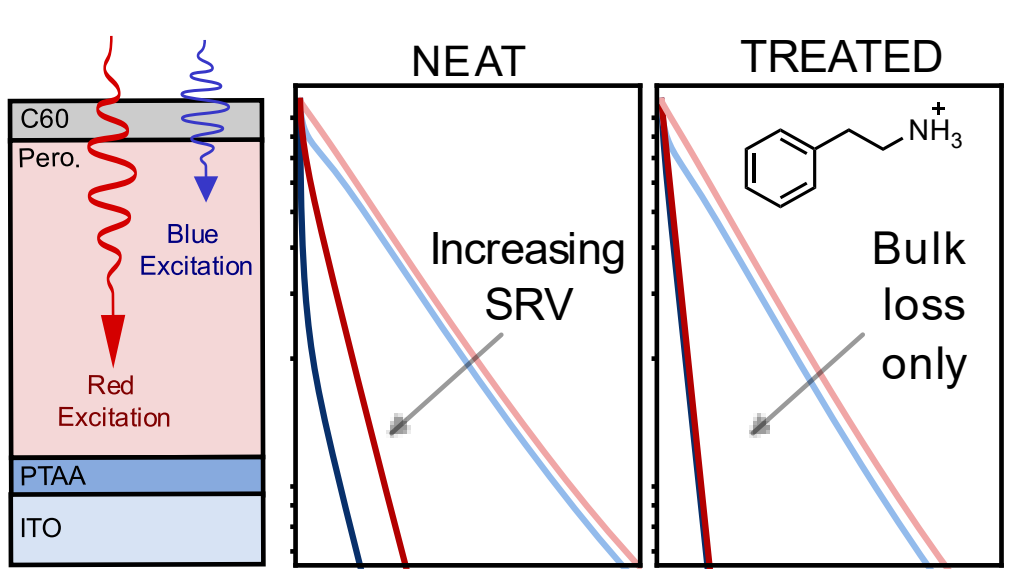

Metal halide perovskites[1] have emerged as an attractive class of semiconductors for applications ranging from photovoltaics,[2] to light emitting diodes (LEDs),[3] sources of quantum light,[4] and beyond.[5] Perovskite solar cells have developed rapidly, with single-junction power conversion efficiency (PCE) reaching 26.9%, and silicon–perovskite tandem PCE now exceeding 34%.[6] Perovskite semiconductors have attractive properties for optoelectronics applications including large absorption coefficients,[7] long carrier diffusion lengths,[8] tunable bandgaps,[9] relative defect tolerance,[10] and compatibility with low-cost manufacturing using both solution[11] and vapor[12] deposition methods across length scales.

Although defect tolerant, perovskites are not defect free.[13–16] Passivating interfaces is critical to achieving improved performance in applications including not just solar cells,[17–22] but also LEDs,[23,24] and single photon emitters.[25–27] However, as device performance improves, researchers are increasingly seeking to understand processes that limit lifetime and stability.[28–31] One of the most challenging tasks in this regard is the ability to assess nondestructively the evolution of new non-radiative losses that may appear in the perovskite bulk or at the interface of an adjacent transport layer under operation or stress.

Existing characterization approaches such as X-ray photoelectron spectroscopy (XPS), time-of-flight secondary ion mass spectrometry (ToF-SIMS), and cross-sectional microscopy provide valuable chemical and structural insight into degradation pathways.[32–35]

However, these methods are generally destructive, require extensive sample preparation, and are typically performed under vacuum or ex situ conditions. [36–38] As such, most techniques are limited in their ability to capture the dynamic evolution of performance losses during actual operating conditions. The development of non-destructive methods[39–43] capable of tracking performance changes and local origins in situ or operando, would therefore offer significant advantages for understanding degradation mechanisms, guiding device optimization, and targeting application of more sophisticated analytical methods.

Here, we explore excitation-wavelength-dependent steady state and time-resolved photoluminescence (PL and TRPL) as tools for depth-sensitive comparison of bulk and interfacial nonradiative losses in partial device stacks. By exploiting differences in absorption depth of our film associated with short and long wavelength excitation, we achieve depth sensitivity without requiring destructive partitioning.

As a model system, we investigate partial stacks incorporating the ammonium cation phenethylammonium (PEA), a widely reported surface passivator known to enhance the efficiency of perovskite solar cells by reducing initial interfacial recombination losses induced by $C_{60}$. Despite initial benefit and promised reported stability of PEA-based treatments[44–47], more recent reports indicate significant degradation of PEA-containing devices under prolonged stress at 85°C or above.[48–51] At these elevated temperatures, PEA is reported to undergo deprotonation and interdiffusion. generating amines that react with $FA^+$ in the perovskite lattice, disrupting structural integrity and charge transport at the interface.[50]

Motivated by these reports, we use $PEABF_4$-treated perovskite stacks as a model system to investigate how wavelength-dependent PL and TRPL respond to aging-induced changes in recombination. Our aim is to assess the depth sensitivity of these optical measurements, rather than establish the detailed chemical pathway responsible for degradation in this system.

During thermal aging of PEA-containing partial device stacks, our results suggest wavelength-dependent PL and TRPL changes consistent with increased recombination near the perovskite/$C_{60}$ interface at early times and greater bulk involvement at longer times. We interpret these optical signatures using one-dimensional diffusion–recombination simulations with ex situ characterization. Together, these results demonstrate the utility of multiwavelength excitation PL and TRPL for nondestructive, depth-sensitive comparisons of interfacial and bulk-involved recombination losses during perovskite degradation.

To establish a model system for depth-sensitive recombination analysis, we first examine the performance benefits of a surface-deposited $PEABF_4$ treatment (0.3 mol% relative to Pb) on perovskite solar cell devices.[52] **Figure 1a** plots the J–V curves under simulated AM1.5 illumination for a typical control device and $PEABF_4$–treated device. The **Figure 1a** inset shows a schematic of the full device p-i-n architecture we used in this study (active area 0.0453 $cm^2$) : indium tin oxide (ITO)/poly[bis(4-phenyl)(2,4,6-trimethylphenyl)amine](PTAA)/$(FA_{0.9}Cs_{0.1})PbI_3$/$C_{60}$/bathocuproine (BCP)/ Au. $PEABF_4$-treated devices exhibit improved initial performance in power conversion efficiency (PCE), open-circuit voltage ($V_{oc}$), and fill factor (FF) compared to neat devices. The device trends in **Fig. 1** are consistent with the statistical device set (n=24) with average PCE increasing from 15.4 ± 0.61% to 17.7 ± 0.3%, $V_{oc}$ increasing from 0.997 ± 0.01 V to 1.051 ± 0.003 V, and FF increasing from 69.0 ± 2.1% to 75.3 ± 1.4% (Table S1) suggesting that $PEABF_4$ treatment reduces recombination losses in unaged device stacks, consistent with previous reports.[53–55]

To evaluate how $PEABF_4$ treatment influences recombination losses, we next performed photoluminescence quantum efficiency (PLQE) measurements, excited with a 532 nm continuous wave laser, on neat and $PEABF_4$-treated perovskite partial stacks (ITO/PTAA/Perovskite) deposited on ITO. **Figure 1b** presents the PL spectra of partial stacks for both conditions without (inset) and with $C_{60}$ deposited.

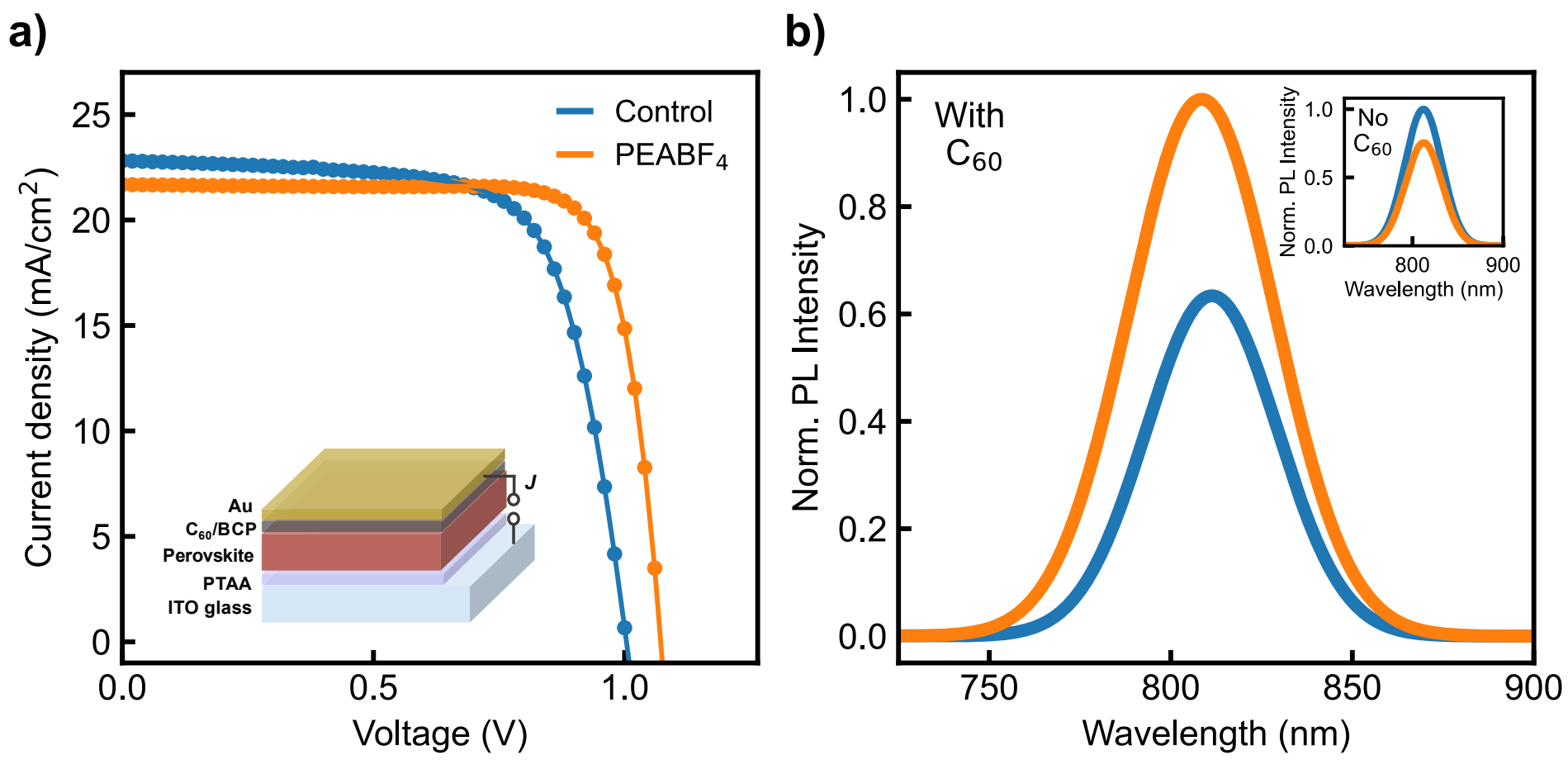

*Figure 1: $PEABF_4$ treatment enhances initial device performance and increases PL emission on partial stacks with fullerene, $C_{60}$, compared to control. (a) Reverse scan J–V curves of representative control (blue) and $PEABF_4$-treated (orange) devices. Schematic of perovskite solar cell structure (p–i–n) with architecture used in this study. (b) Normalized photoluminescence spectra of control (blue) and $PEABF_4$-treated (orange) partial stacks with $C_{60}$. Inset shows normalized photoluminescence spectra of partial stacks without $C_{60}$.*

Despite many reports on the PL benefit of surface-deposited PEA-containing treatments on bare film,[44,52,56,57] we observe a decrease in PL compared to neat film. To determine whether this response arises from the $PEA^+$ concentration alone, we compared concentration-dependent PLQE trends for $PEABF_4$- and PEAI-treated partial stacks (Figure S1). Increasing the $PEABF_4$ concentration does not produce a PLQE enhancement, whereas PEAI treatment yields increased PLQE at intermediate concentrations. These contrasting trends indicate that the PL response cannot be attributed solely to the presence or concentration of $PEA^+$ and instead depends on the accompanying anion and resulting surface chemistry. Thus, under our processing conditions, $PEABF_4$ treatment does not enhance the PL of the perovskite before $C_{60}$ deposition.

However, after $C_{60}$ deposition, the $PEABF_4$-treated films show over a 9-fold higher PLQE than the $C_{60}$-contacted control films, increasing from $0.061 \pm 0.002$ % to $0.564 \pm 0.041$%. Thus, in our stacks, the beneficial effect of $PEABF_4$ emerges specifically after formation of the $C_{60}$/perovskite interface where the treatment appears to mitigate part of the PLQE loss associated with $C_{60}$ deposition. Based on the improvement in PL in the $PEABF_4$-treated $C_{60}$-contacted films relative to those without $PEABF_4$ treatment, we estimate that the QFLS of film with $PEABF_4$ increases by $57 \pm 3$ meV. This value agrees with the device-measured $V_{oc}$ increase of $54 \pm 6$ mV, suggesting that the measured gain in $V_{oc}$ is consistent with the improvement in the PL-derived internal voltage limit.

The fact that $PEABF_4$ only improves the relative PLQE of the partial stacks after $C_{60}$ addition strongly suggests an interfacial effect at that interface, providing a useful model system to explore depth-sensitive comparison of recombination losses via excitation-wavelength-dependent PL measurements. To test this hypothesis, we next measured excitation-wavelength-dependent wide-field PL excited with collimated LEDs with peak wavelengths of 450 nm (blue) and 660 nm (red) excitation. Based on ellipsometry-derived absorption coefficients (Table S2), the corresponding optical penetration depths (1/e decay depth) in our perovskite layers are approximately 55 nm (450 nm excitation) and 337 nm (660 nm excitation) (Figure S2). Therefore, we expect PL detected after blue excitation to be qualitatively more representative of recombination in near-surface regions and red-excited PL to be more qualitatively sensitive to excitations deeper in the film. To selectively interrogate the perovskite/$C_{60}$ interface where $PEABF_4$ and $C_{60}$ are introduced, we illuminated the partial stacks from the top perovskite/ETL side and varied the excitation fluence over nearly three orders of magnitude (Figure S3) to control for any variations in recombination order.

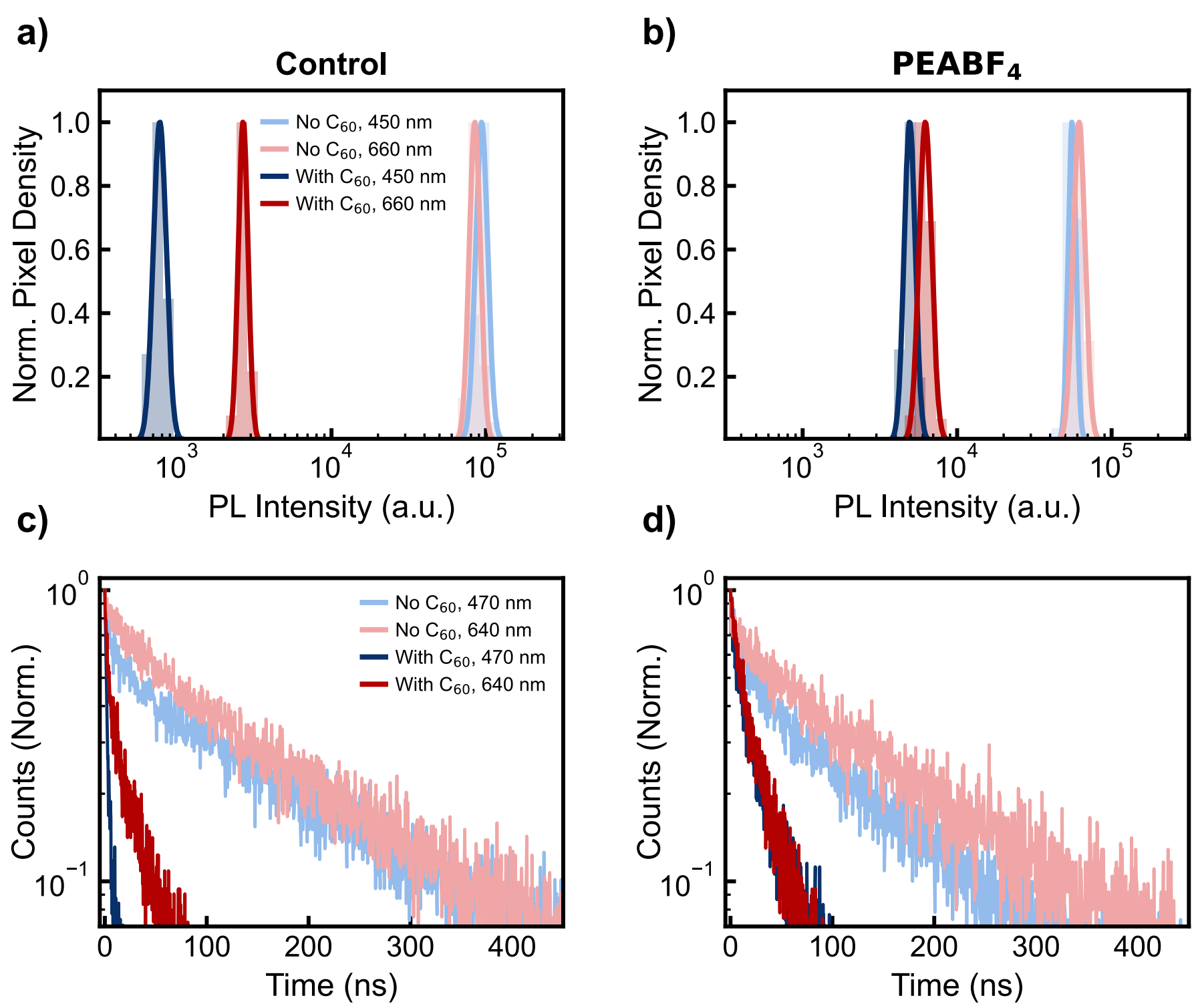

***Figure 2***. Excitation-wavelength-dependent *photoluminescence shows $PEABF_4$ mitigates* $C_{60}$ */perovskite interfacial recombination. Steady state PL measurements of partial stacks on ITO for (a) control and (b) $PEABF_4$-incorporated samples without (lighter colors) and with* $C_{60}$ (darker colors), *illuminated at approximately 1 sun illumination. 450 nm and 660 nm collimated LEDs used as blue and red excitation probes and matched at 1.6e17 photons $cm^{-2}$ $s^{-1}$. (c) Transient decays of control and (d) PEA-treated samples.* $C_{60}$*-incorporated decays indicated. 470nm and*

*640nm pulsed lasers used as blue and red excitation probes and matched at 1.6e10 photons cm$^{-2}$ pulse$^{-1}$ .*

**Figure 2a** shows the PL intensity distributions from images of control samples with and without $C_{60}$ ETL layers. Depending on excitation wavelength, the sample without $C_{60}$ shows ≥30-fold higher PL than the sample with $C_{60}$. These results are consistent with the quantitative PLQE measurements discussed above. In addition, without $C_{60}$ the PL intensity distributions overlap when excited with either red or blue light. Notably, however, after $C_{60}$ deposition, the PL is dramatically lower when excited with blue light relative to the PL when excited with red light. We interpret this difference as resulting from the high surface-recombination velocity at an unpassivated $C_{60}$/perovskite interface inducing fast non-radiative losses for carriers near that interface, consistent with a number of prior reports.[58,59]

For comparison, **Figure 2b** shows the steady-state PL intensity distributions from images of $PEABF_4$-treated samples with and without $C_{60}$ ETL layers. Samples without $C_{60}$ show ~10X higher PL intensity than samples with $C_{60}$ (this is a smaller level of quenching than $C_{60}$ induces in the control samples). As before, without $C_{60}$, the PL intensity distributions overlap when excited with either red or blue light, with the blue-excited PL being only slightly lower. After $C_{60}$ deposition, the red-blue PL difference in the $PEABF_4$-treated films is much smaller than that observed for the control films in **Figure 2a**. Figure S4 shows representative steady state widefield PL images of untreated control and $PEABF_4$-treated samples before and after $C_{60}$ deposition.

Together we interpret the results in **Figures 2a** and **2b** as follows. First, the $C_{60}$/perovskite contact is one that supports a high rate of interfacial recombination, with surface recombination velocities (SRVs) on the order of a few thousand cm s$^{-1}$ commonly reported.[58,59] Second, $PEABF_4$ deposition strongly suppresses the SRV at the perovskite/$C_{60}$ interface. This suppression could result from field effect passivation,[60,61] or simple physical decoupling of the perovskite and $C_{60}$.[62] These two effects qualitatively explain the most significant trends in the data, strong PL quenching consistent with enhanced recombination near the perovskite/ETL interface (blue excited PL quenched more than red-excited PL) following deposition of $C_{60}$, along with strong reduction in these trends when $PEABF_4$ surface treatment is present. Finally, to explain the remaining reduced, and comparatively equal quenching of blue- and red-excited PL that remains with $PEABF_4$ surface following $C_{60}$ deposition, we speculate that either our process for the vacuum thermal evaporation of $C_{60}$ leads to some damage to the perovskite film, resulting in an increase in bulk defect densities, or that there is some degree of surface roughness and associated exposed lateral interfaces (grain boundaries) that provide more uniform quenching across the film thickness and may be partially infiltrated by fullerene.

Next, we explore these hypotheses in more detail with TRPL and simulations. **Figure 2c** shows the excitation-wavelength-dependent TRPL data on the same control samples shown in **Figure 2a**. Consistent with the steady state PL image data, the TRPL lifetimes of the control sample without the $C_{60}$ ETL are nearly identical when excited with red or blue light. However, upon deposition of the $C_{60}$ layer, the TRPL lifetimes drop dramatically, with the blue-excited TRPL dropping the most, consistent with our interpretation of the decrease in blue-PL as being due to increased non-radiative recombination from carriers excited near the perovskite/ETL interface (see below).

Likewise, **Figure 2d** displays the wavelength-dependent TRPL data on the $PEABF_4$-treated samples shown in **Figure 2b**. Similar to the steady state PL results, samples without $C_{60}$ show largely overlapping TRPL decays under red and blue excitation, with the blue-excited decay being slightly faster. With $C_{60}$ deposition, both the red- and blue-excited decays shorten overall. However, the two excitation wavelengths remain overlapping, indicating that $C_{60}$ increases recombination losses without producing a strong wavelength-dependent difference in the recombination dynamics.

These TRPL results are also in good qualitative agreement with the interpretation proposed above for the data in Figures 2a and 2b. Next, we turn to ground this conversation in a quantitative analysis. To do so, we use a one-dimensional diffusion–recombination model that accounts for bulk recombination lifetime ($\tau_{bulk}$), carrier diffusion, shallow trapping and detrapping, and interfacial recombination through $SRV$ boundary conditions (Supporting Section I). We initialized our model using experimental pulse fluence and wavelength-dependent absorption coefficients for a 600 nm-thick film. We then varied mobility ($\mu$), $\tau_{bulk}$, and HTL- or ETL-side surface recombination velocities ($SRV_{HTL}$ & $SRV_{ETL}$) within literature-relevant ranges in Figure S5.[63–66]

We emphasize that these simulations are not intended to uniquely extract fitted parameters, but to help determine if our results and interpretation are consistent with the experimentally observed excitation-wavelength dependent behavior. Specifically, we ask whether it is possible to quantitatively reproduce the observed trends with physically plausible values of the bulk and interfacial recombination parameters in the presence and absence of $C_{60}$ and $PEABF_4$ and combinations thereof.

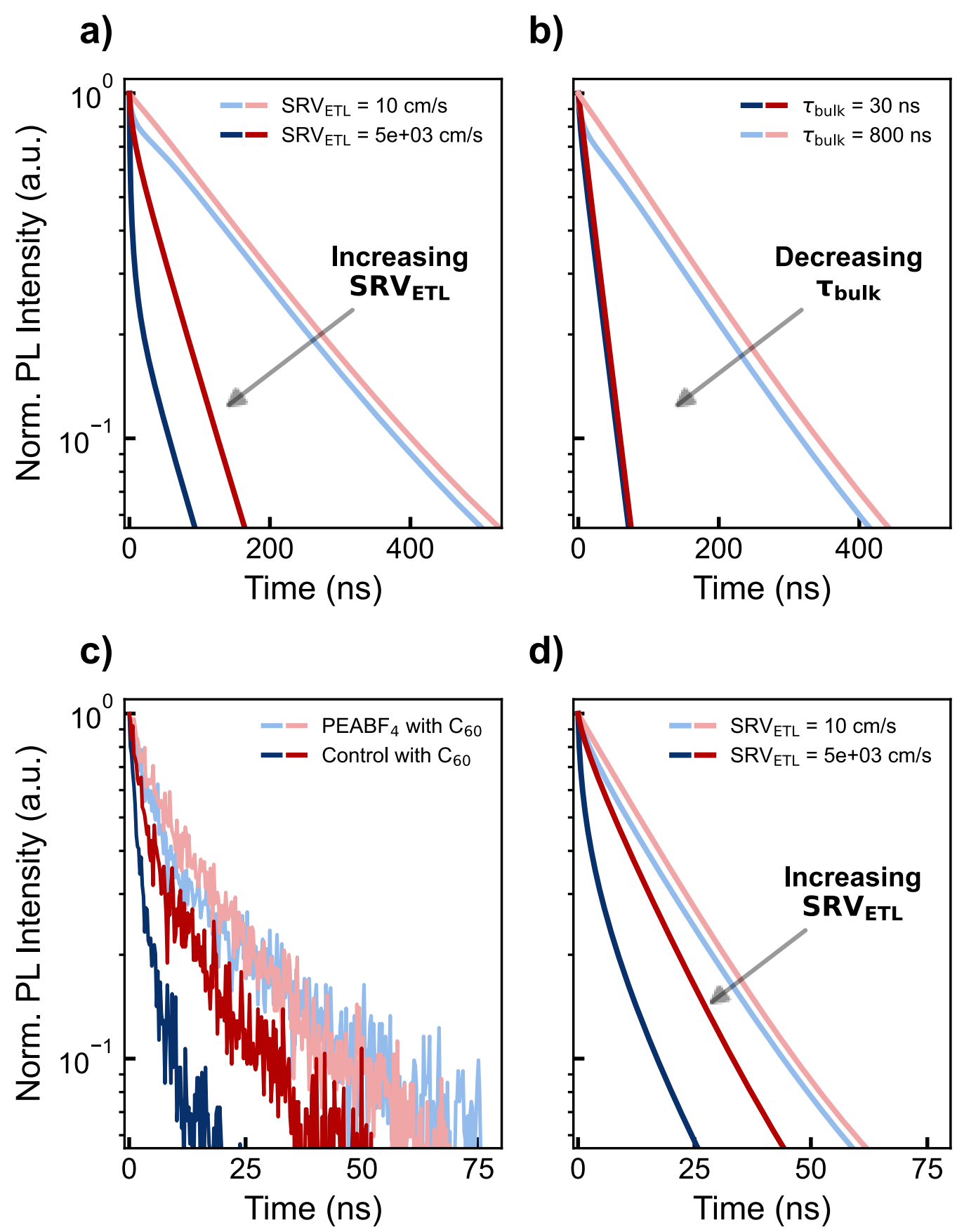

***Figure 3**. Simulated decrease in ETL-side surface recombination velocity show differentiation between excitation-wavelength-dependent decays while bulk lifetime decrease reveals wavelength-independent fast decays that approach experimental decays for films containing* $C_{60}$*. Sensitivity of blue- and red-excited TRPL decays for partial stacks, simulated with (a)* $SRV_{ETL}$ *= 10 cm* $s^{-1}$ *(lighter colors) or* $SRV_{ETL}$ *= 5000 cm* $s^{-1}$ *(darker colors) with* $\tau_{bulk}$ *= 800 ns and (b)* $\tau_{bulk}$ *= 800 ns (lighter colors) or* $\tau_{bulk}$ *= 30 ns (darker colors), with* $SRV_{ETL}$ *= 50cm* $s^{-1}$*. (c) Experimental decays for neat and* $PEABF_4$*-treated samples with* $C_{60}$ *and (d) simulation carried out with* $SRV_{ETL}$ *= 10 cm* $s^{-1}$ *or* $SRV_{ETL}$ *= 5000 cm* $s^{-1}$ *,with* $\tau_{bulk}$ *= 30ns and all other parameters unchanged. Simulated decays for all swept-through values were overlayed on top of the representative experimental decays and are shown in Figure S6. All additional parameters for each simulation are reported in Table S3.*

SRV is commonly used to describe carrier loss at the semiconductor surface or interface and parameterize nonradiative interfacial recombination.[58,63,67–69] Because $C_{60}$ shows such strong interfacial quenching, we assign $SRV_{ETL}$ as the primary model parameter to represent enhanced carrier loss at the perovskite/$C_{60}$ interface.

**Figure 3a,b** displays simulations that explore the key experimental trends discussed above. First, **Figure 3a** shows simulated blue- and red-excited TRPL decays with $SRV_{ETL}$values set at either 5000 cm $s^{-1}$ (with the ranges reported for $C_{60}$-contacted films) or 10 cm $s^{-1}$ (within the range of well-passivated perovskite interfaces)[58,59,61,65] and $\tau_{bulk}$ set to 800ns. As anticipated, with ETL-side illumination, we observe that higher values of $SRV_{ETL}$ lead to overall faster decays with distinct excitation-wavelength-dependent separation: blue-excited decays in particular become faster compared to red-excited decays. This observation is consistent with the experimental trend in **Figure 2c** which compares neat samples with and without $C_{60}$. As expected, our transient simulations suggest that, under ETL-side illumination, inducing fast non-radiative losses for carriers near the ETL interface show overall reduced TRPL lifetimes with losses at the surface (blue- excited) notably displaying large reductions.

**Figure 3b** shows simulated TRPL decays set with $\tau_{bulk}$ values at either 800ns or 30ns. Indeed, we observe that lower $\tau_{bulk}$ values shorten decays overall without showing contrast in blue- vs. red-excited TRPL decays (i.e., more vertically homogeneous recombination) as indicated by the overlap of blue- and red- excited decays. Interestingly, we identify that these simulated decays align with experimental $PEABF_4$-treated decays with and without $C_{60}$ (**Figure 2d**). These simulations suggest that our fullerene processing or infiltration along grain boundaries can possibly introduce near-surface or bulk-like recombination losses as proposed by early work.[70,71]

As a cross check on the use of such fast bulk lifetimes, we ran steady state simulations to investigate whether this low parameter would still achieve a sensible $V_{oc}$ value. We found that the resulting voltages are in the range of our experimental values (Figure S7a). We additionally have demonstrated relative steady state PL emission simulations that show at higher $SRV$ values we achieve trends on par with $C_{60}$ deposition where excited depths further from the largely quenching ETL show higher PL (Figure S7b). To experimentally validate the illumination-direction dependence predicted by the steady state PL emission simulations, we performed steady state measurements on partial stacks using excitation from both the ETL ($C_{60}$) side and HTL (PTAA) side, both before and after $C_{60}$ deposition as shown in Figure S8.

Based on the simulation trends described above, larger $SRV_{ETL}$values are expected to produce stronger vertical contrast between blue- and red-excited decays, whereas shorter $\tau_{bulk}$ values should accelerate both decays without introducing strong wavelength-dependent differences. We therefore compared these trends to the measured TRPL decays of neat and $PEABF_4$-passivated samples with $C_{60}$, as shown in **Figure 3c**. **Figure 3d** shows simulated transient decays in which we set $\tau_{bulk}$ to be small (30ns) to reflect the fast, homogeneous decays of $PEABF_4$-passivated samples with $C_{60}$. We then varied $SRV_{ETL}$ values to either 5000 cm $s^{-1}$ or 10 cm $s^{-1}$ to model the effect of interfacial recombination. As expected, increasing $SRV_{ETL}$produces stronger separation between the red- and blue-excited decays, consistent with the greater excitation-wavelength-dependent contrast observed experimentally in the neat samples compared to the $PEABF_4$-treated samples. We report all additional simulations with their parameters in Supporting Information Section I.

Finally, having analyzed the effects of $PEABF_4$ and described the variations in PL intensity and lifetime with different treatments, we next turn to examine the effects of thermal stress on our samples with and without $PEABF_4$. Following the aging protocol reported by Park et al.[72], we subjected devices to thermal aging by placing them on a hotplate at 85°C in a nitrogen-filled glovebox under dark conditions. We then measured device properties periodically for up to 120 hours (Figure S11). **Figure 4a,b** plots the J–V curves under simulated AM1.5 illumination for a typical control device and PEA-treated device recorded at 0, 2, 24 and 48 hours of thermal aging at 85°C.

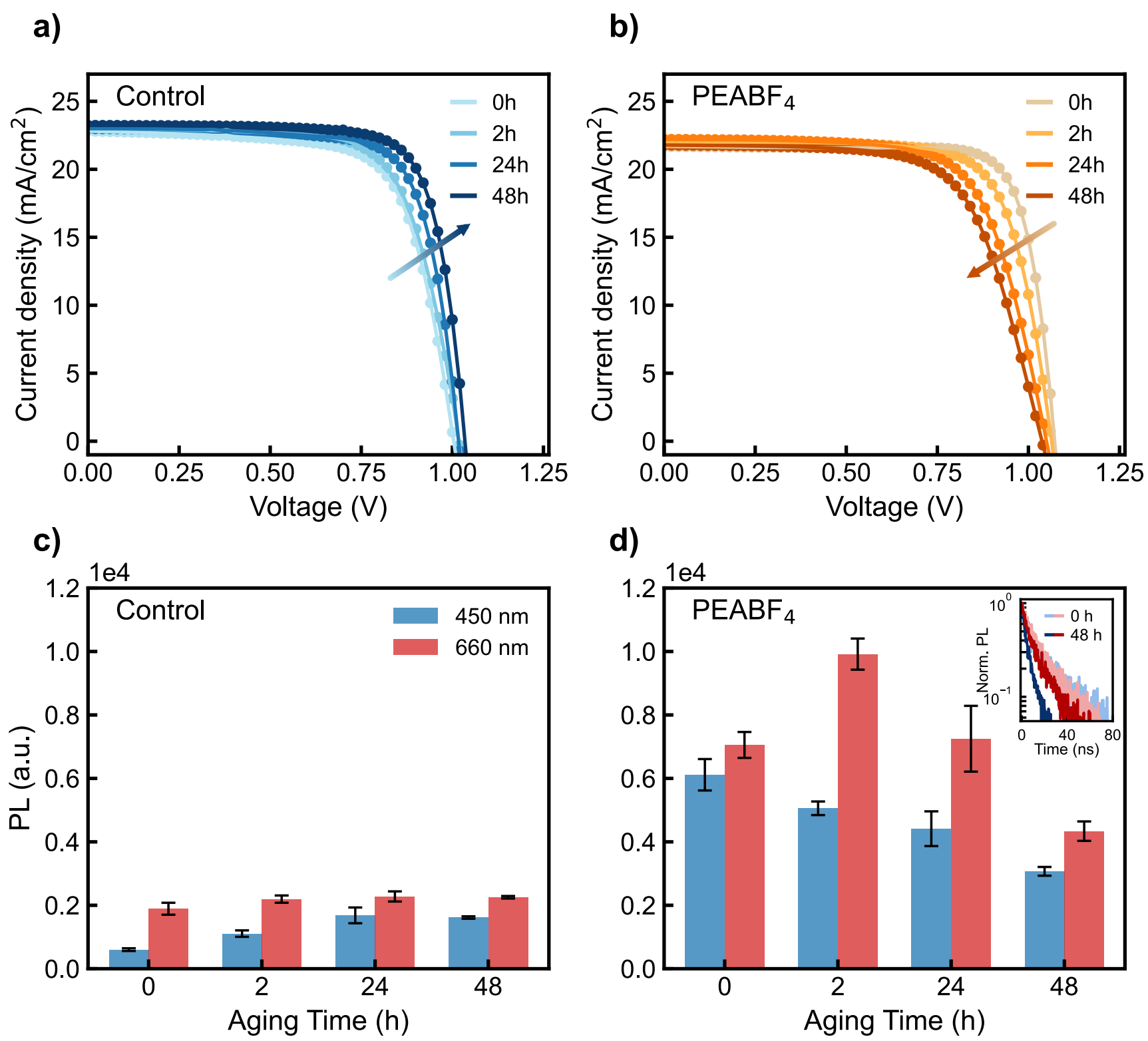

*Figure 4. $PEABF_4$ treatment enhances initial performance, but decreases thermal stability compared to control. (a) J–V curves of representative control perovskite solar cell subject to continuous thermal aging at 85°C (b) J–V curves of continuously aged $PEABF_4$-treated solar cells. Average red- and blue- excited PL emission trends of (c) control and (d) $PEABF_4$-treated partial stacks with $C_{60}$ thermally aged over 48 hours. Inset shows representative decays for $PEABF_4$-treated samples with $C_{60}$ before and after 48 hours of aging.*

After 2 hours of thermal aging, the typical control and $PEABF_4$-treated devices exhibit slight improvements in $V_{oc}$ consistent with the statistical device set (19 ± 6.4 mV increase for neat and 9 ± 3.2 mV increase for $PEABF_4$). This improvement in J–V characteristics at early times is known from the literature has been previously attributed to ion redistribution and improved compositional homogeneity.[73–75]

However, at extended thermal aging we find distinct behaviors emerge between the control and $PEABF_4$-treated films. **Figure 4a** shows that the J–V sweeps of the control devices improve slightly and gradually with thermal aging. In contrast, **Figure 4b** shows that $PEABF_4$-treated devices display progressive degradation with noticeable reductions in device parameters over time (see Table S4 and Table S5 for full statistics across n=24 aged devices).

To probe changes in the PEA distribution during thermal aging, we performed depth-dependent time-of-flight secondary ion mass spectrometry (ToF-SIMS) depth-profile measurements on $PEABF_4$-treated partial stacks after 0, 2, 24, and 48 h of aging at 85 °C as shown in Figure S12a. Relative to the unaged film, the aged profiles show changes in PEA-associated secondary-ion signals at longer sputter times, consistent with redistribution of PEA-derived species. To characterize these changes, we calculated the ratio of the mean PEA-associated ion fraction in selected bulk and surface windows. This bulk-to-surface signal ratio increases from 0.22 in the unaged film to 10.43 after 2 h and remains elevated after 24 and 48 h (Figure S12b). Thus, an altered PEA-associated depth distribution is already evident after 2 h and persists during the more pronounced optoelectronic degradation observed at 24–48 h. However, persistence of this redistributed profile does not establish that continued redistribution drives the later increase in recombination losses. These measurements therefore provide complementary chemical context for the optical results without identifying the chemical form of the redistributed species or establishing the mechanism responsible for the evolving recombination losses.

To study the concomitant evolution of nonradiative losses with thermal aging, we performed subsequent measurements on partial stacks with $C_{60}$ and subjected them to 0, 2, 24, 48 hours of continuous thermal aging at 85°C. We first measured PLQE measurements on (n = 4) films per condition and plot the average calculated QFLS values for aged neat and $PEABF_4$-treated sample stacks in Figure S14. We find these QFLS trends agree with the corresponding device-measured $V_{oc}$ evolution. With aging, both metrics show that the initial advantage of $PEABF_4$-treatment is maintained at 2 h, largely lost by 24 h, and further diminished by 48 h where QFLS decreases by 73 meV and $V_{oc}$ decreases by 66 mV from 0 to 48 h. In contrast, the control shows a comparably constant QFLS over aging or improvement in $V_{oc}$. Table S6 and Table S7 show a full summary of $V_{oc}$ and QFLS changes with aging.

To determine whether these aging-induced losses are more strongly associated with the $C_{60}$-contracted interface or extend into the perovskite bulk, we used multi-wavelength excitation widefield steady-state PL and TRPL characterization over multiple fluences in samples with $C_{60}$ (Figure S15 & S16, Table S8 & Table S9). **Figure 4c,d** shows the average red- and blue-excited PL emission for neat and $PEABF_4$-treated stacks as a function of thermal stressing time when excited with 1-sun intensity. In control samples, we observe relatively consistent PL emission over the course of thermal stress history, with only minor increases in PL in blue-excited emission. In contrast, $PEABF_4$-treated samples appear to show an evolution in the trends of blue-excited and red-excited PL emission. After 2 hours of aging, we find that $PEABF_4$-treated samples display a localized decrease in PL emission with blue-excitation, while after longer aging times of 24–48 hours, we observe suppression of both blue- and red-excited PL. We interpret this gradual evolution of recombination loss to signify that early degradation is initially strongly associated with the perovskite/$C_{60}$ interface and later extends into a larger fraction of the perovskite bulk when annealing at longer times in the presence of $PEABF_4$. Figure S17 and Figure S18 shows representative widefield PL images for aged neat and $PEABF_4$-incorporated stacks.

Finally, to test the $PEABF_4$-induced degradation trends measured from steady-state PL, we examine blue- and red-excited TRPL decays. Figure S19 shows representative blue- and red-excited TRPL decays for $PEABF_4$-treated samples before and after continuous thermal aging. Like our steady-state emission trends, $PEABF_4$-treated samples show steady shortening of more interface-sensitive decays (blue), while decrease of less surface-sensitive decays (red) isn't seen until after longer aging times (24 hours).

Figure 4d inset shows typical decays for $PEABF_4$-treated samples with $C_{60}$ before and after 48 hours of aging for both excitation-wavelengths. Before aging, $PEABF_4$-treated samples with $C_{60}$ show overlapping red- and blue- excited decays. However, after 48 hours of aging, decays shorten overall where the blue-excited TRPL decay shows greater reduction compared to the red-excited TRPL decay. This trend is similar to decays comparing unaged control and $PEABF_4$-containing samples with $C_{60}$ as shown in **Fig. 4c,d**. We interpret this similarity between the two $C_{60}$-containing sample trends to the combined effect of lower $\tau_{bulk}$ ($C_{60}$-induced) and increasing $SRV_{ETL}$ values (more interfacial recombination). Section I, Figure S9 plots the comparison between the two decay trends.

Together, these results support our interpretation that while $PEABF_4$ treatment initially reduces nonradiative-recombination near the perovskite/$C_{60}$ interface, thermally induced instability results in gradual recombination losses that evolve and extend from the perovskite/$C_{60}$ interface into the bulk. We provide comparatively stable aged neat sample decays in Figure S20. Measurements using additional excitation wavelengths, illumination from both sides of the stack, and a range of excitation fluences would provide complementary steady-state intensities and transient decays that could better constrain a semi-quantitative depth-dependent model.

We demonstrate that excitation-wavelength-dependent steady state and time-resolved photoluminescence are powerful tools that allow us to nondestructively infer evolution of recombination losses in perovskite partial device stacks. Using $PEABF_4$-treated perovskite stacks as a model system, we find that the treatment provides little benefit to bare perovskites, but in device stacks it initially suppresses $C_{60}$-associated interfacial recombination. Thermal aging produces depth-sensitive PL/TRPL signatures consistent with recombination losses that initiate near the $C_{60}$-contacted interface and later affect a larger fraction of the absorber. Multiwavelength excitation-dependent PL/TRPL therefore provides a nondestructive screening approach for evaluating whether aging-induced losses remain interface-localized or become spatially distributed. We anticipate that these techniques can serve as a valuable platform for guiding the optimization of interfaces and passivation strategies in future device development.

**Supporting Information**

Concentration-dependent PLQE measurements; ellipsometry-derived absorption coefficients and penetration depths; excitation-wavelength-, fluence-, and illumination-side-dependent steady-state PL measurements; widefield PL images; simulated PL and TRPL responses to transport, recombination, and trapping parameters; experimental and simulated TRPL comparisons; photovoltaic characteristics during thermal aging; ToF-SIMS depth profiles of PEA- and FA-associated species; aging-dependent PLQE and quasi-Fermi-level splitting measurements; and stretched-exponential TRPL fits (PDF).

**Acknowledgements**

This material is based upon work that began with funding by the U.S. Department of Energy's Office of Energy Efficiency and Renewable Energy (EERE) under the Solar Energy Technologies Office Award Number DE-EE0011422, but was primarily supported by J.J.'s work under a National Science Foundation Graduate Research Fellowship under Grant No. DGE-2140004. The views expressed herein do not necessarily represent the views of the U.S. Government or any agency thereof.

The authors acknowledge the use of facilities and instruments at the Photonics Research Center (PRC) at the Department of Chemistry, University of Washington; Research Training Testbed (RTT), part of the Washington Clean Energy Testbeds system. Part of this work was conducted at the Molecular Analysis Facility (MAF), a National Nanotechnology Coordinated Infrastructure site at the University of Washington which is supported in part by funds from the National Science Foundation (awards NNCI-2025489, NNCI-1542101), the Molecular Engineering & Sciences Institute. D.S.G. and J.J acknowledge support from the Washington Research Foundation and Rabinovitch Endowment for completing the manuscript following the termination of the award.

Supporting Information: Additional experimental details, materials, and methods (PDF).